\documentclass[11pt, twocolumn, twoside, a4paper]{article}

\usepackage[top=2cm, bottom=2cm, left=1.5cm, right=1.5cm]{geometry}
\usepackage{authblk}

\usepackage[T1]{fontenc}
\usepackage{amssymb}
\usepackage{amsmath}

\usepackage{graphicx}
\usepackage{caption}
\usepackage{float}

\usepackage{siunitx}
\DeclareSIUnit[number-unit-product = {}]
\uL{\micro\liter}
\DeclareSIUnit[number-unit-product = {}]
\umol{\micro \mole}
\DeclareSIUnit[number-unit-product = {}]
\uM{\micro M}

\newcommand{\textsub}[1]{$_{\text{#1}}$}

\title{Lipid membrane-mediated attraction between curvature inducing objects}

\author[1]{Casper~van~der~Wel}
\author[2]{Afshin~Vahid}
\author[3]{An\dj ela~\v{S}ari\'c}
\author[2]{Timon~Idema}
\author[1, 4]{Doris~Heinrich}
\author[1, *]{Daniela~J.~Kraft}

\affil[1]{Biological and Soft Matter Physics, Huygens-Kamerlingh Onnes Laboratory, Leiden University, PO Box 9504, 2300 RA Leiden, The Netherlands}
\affil[2]{Department of Bionanoscience, Kavli Institute of Nanoscience, Delft University of Technology, Van der Maasweg 9, 2629 HZ Delft, The Netherlands}
\affil[3]{Department of  Physics and Astronomy, Institute for the Physics of Living Systems, University College London, Gower Street, London, WC1E 6BT}
\affil[4]{Fraunhofer Institute for Silicate Research, Neunerplatz 2, 97082 W\"urzburg, Germany}
\affil[*]{kraft@physics.leidenuniv.nl}

\date{}

\begin{document}
\maketitle

\begin{abstract}  
	The interplay of membrane proteins is vital for many biological processes, such as cellular transport, cell division, and signal transduction between nerve cells.
	Theoretical considerations have led to the idea that the membrane itself mediates protein self-organization in these processes through minimization of membrane curvature energy. 
	Here, we present a combined experimental and numerical study in which we quantify these interactions directly for the first time.
	In our experimental model system we control the deformation of a lipid membrane by adhering colloidal particles.
	Using confocal microscopy, we establish that these membrane deformations cause an attractive interaction force leading to reversible binding.
	The attraction extends over 2.5 times the particle diameter and has a strength of three times the thermal energy (\SI{-3.3}{k\textsub{B}T}).
	Coarse-grained Monte-Carlo simulations of the system are in excellent agreement with the experimental results and prove that the measured interaction is independent of length scale.
	Our combined experimental and numerical results reveal membrane curvature as a common  physical origin for interactions between any membrane-deforming objects, from nanometre-sized proteins to micrometre-sized particles.
\end{abstract}

\section*{Introduction}
Interactions between membrane proteins are of key importance for the survival of cells as they are involved in many dynamical processes.
The organization of membrane proteins into complexes and their effect on membrane shape enables for instance intracellular transport, cell division, cell migration, and signal transduction \cite{McMahon2005}.
Understanding the underlying principles of protein organization is therefore crucial to unravel processes such as cell-cell signalling in the brain \cite{Freche2011} or disease mechanisms like membrane-associated protein aggregation in Parkinson's disease \cite{Pfefferkorn2012}.

Besides specific protein-protein interactions and interactions with the cytoskeleton, protein organization in membranes is thought to be driven by a universal interaction force arising from membrane deformations.
Theoretical models \cite{Goulian1993a,Kim1998,Dommersnes2002,Muller2005,Yolcu2014} and simulations \cite{Reynwar2007,Saric2013a,Simunovic2013} predict that by deforming the membrane locally, membrane proteins can self-assemble into complex structures such as lines, rings, and ordered packings \cite{Saric2013a,Pamies2011, Saric2012a, Vahid2015}.
Observations in living cells \cite{Pfefferkorn2012, Peter2004a} support the existence of such membrane-mediated interactions, but have yet to provide conclusive experimental proof of their common physical origin: separation of contributions arising from specific protein-protein interactions and interactions with the cytoskeleton is extremely challenging.

Further experimental indications for a universal membrane-mediated interaction stem from simplified model systems: phase-separated membrane domains are known to repel each other \cite{Semrau2009} while colloidal particles have been observed to irreversibly stick together when attached to lipid vesicles \cite{Koltover1999, Ramos1999}.
However, the hypothesized connection between curvature and interaction force has not been quantified to date: even the sign of the force is still under debate.

Existing model systems for studying surface-mediated interactions are typically based on deformations of liquid-liquid or liquid-air interfaces \cite{Loudet2005, Cavallaro2011, Vella2005}.
In these systems, interactions are governed by surface tension, while in lipid vesicles elastic \emph{surface bending} is expected to be the dominant factor.
In addition, lipid vesicles are bilayers of molecules that cannot exchange molecules with the surrounding medium, which makes them profoundly different from other liquid interfaces.
The experimental quantification of interface-mediated interactions in lipid membranes thus requires a clean and dedicated model system.

\begin{figure}[ht]  
	\centering{\includegraphics{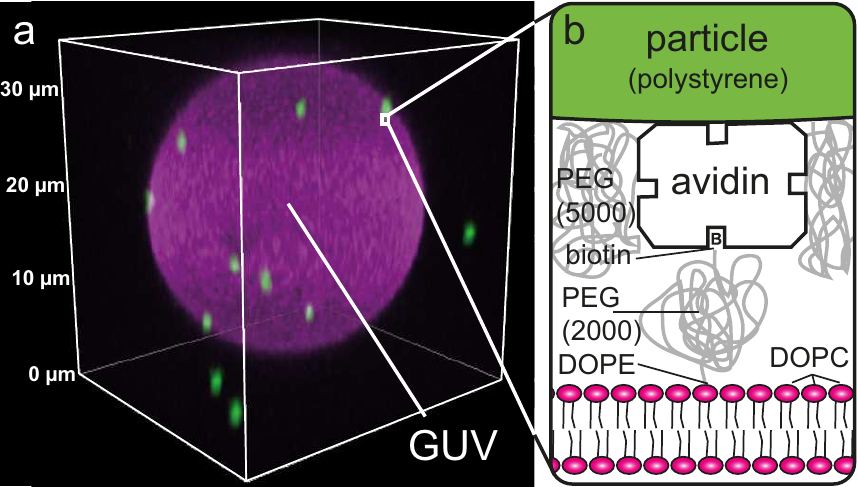}}
	\caption{Experimental model system for studying membrane-mediated interactions.
		(a) Three-dimensional confocal image of a typical Giant Unilamellar Vesicle (GUV, in magenta) with attached colloidal particles (in green). Supplementary Video S1 contains the corresponding movie.
		(b) Schematic of the avidin-biotin linkage between membrane and particle.
		By varying the avidin concentration on the particles we control the adhesion strength.
		Polyethylene glycol (PEG) suppresses electrostatic interactions between membrane and particles, as well as non-specific adhesion between particles.}
	\label{fig:sketch}
\end{figure}

In this article, we describe such a specialized model system consisting of membrane-adhering colloidal particles on Giant Unilamellar Vesicles (GUVs).
We characterize for the first time the effect of a single adhesive colloidal particle on the local membrane shape using confocal microscopy.
We find that the particle is either fully wrapped by the membrane or not wrapped at all, depending on the adhesion strength.
Next, we measure the interaction potential for particles in these two states and we find that only wrapped particles show a reversible attraction, which implies that the attraction is purely caused by the membrane deformation.
Monte Carlo simulations of the bending-mediated interaction between wrapped particles result in an interaction potential that quantitatively agrees with the experimental result.
Since these simulations do not contain any absolute length scale, we conclude that the measured attraction caused by lipid membrane deformations is scale-independent. Our combined model system and simulations therefore quantitatively describe the interactions of any membrane-deforming object, ranging from nanometre-sized proteins to micrometre-sized colloidal particles.

\section*{Results}
\subsection*{Particle-induced membrane deformation}
As a dedicated model system for membrane-deforming proteins we use micrometre-sized colloidal particles (poly\-styrene, \SI{0.98+-0.03}{\um} in diameter) adhered to single-component Giant Unilamellar Vesicles (GUVs, diameters ranging from \SIrange{5}{100}{\um}), allowing us to study membrane-mediated interactions with confocal microscopy (see Figure \ref{fig:sketch}).
The GUVs consists of DOPC lipids, which is above its melting point at room temperature, ensuring a single-phase liquid membrane.
The connection between membrane and particle is realized by coating the particles with varying amounts of avidin, a protein that binds strongly and specifically to biotin \cite{Moy1994}, which we attach to the membrane through a functionalised lipid.
The concentration of avidin linkers on the particle surface allows us to effectively tune the adhesion strength of the particle to the membrane.

\begin{figure}[h]
	\centering{\includegraphics{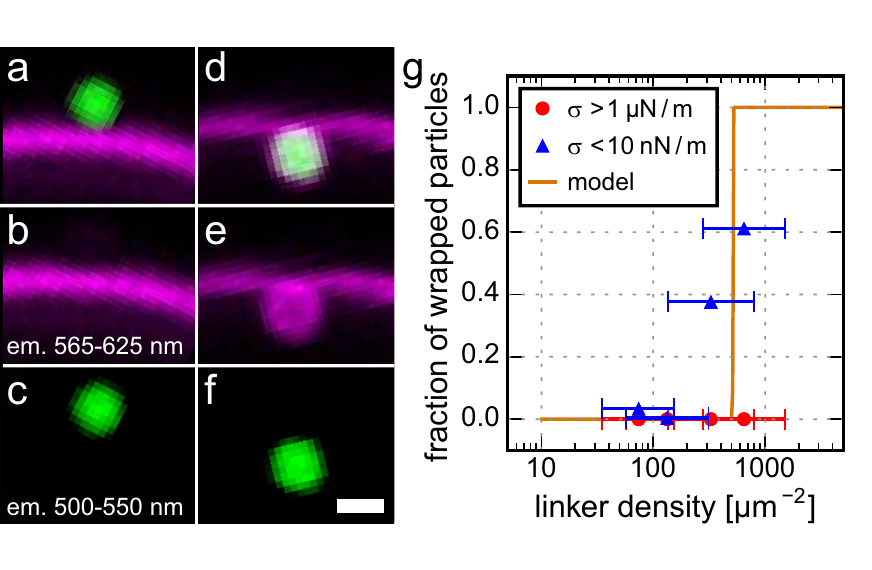}}
	\caption
	{The effect of particle linker density on the membrane wrapping state.
		(a) Fluorescence signal of a non-wrapped particle (green) and a membrane (magenta).
		The separate fluorescence signals of the membrane and particle are displayed in (b) and (c), respectively.
		In (d)-(f) the wrapped state is displayed analogously.
		The scale bar is \SI{1}{\um}.
		(g) Fraction of wrapped particles as a function of linker density on floppy membranes (blue triangles, membrane tension $\sigma < \SI{10}{nN \per \meter}$) and tense membranes (red circles, $\sigma > \SI{1}{\micro N \per \meter}$). The solid line is the analytic model at $\sigma = 0$ derived from Equation \ref{eq:E} via the Boltzmann factor.
		Horizontal error bars show the spread (one standard deviation) in linker density.}
	\label{fig:wrapping}
\end{figure}

By choosing different fluorescent markers for the particles and lipid membranes, we are able to visualize the effect of a single particle on a lipid membrane (see Figure \ref{fig:wrapping}).
We find that particles exist in either a completely wrapped state or a completely non-wrapped state: partial wrapping is only observed as a transient situation.
Non-wrapped particles are located on the outside of the vesicle without deforming the membrane (Figure \ref{fig:wrapping}a-c), while wrapped particles are protruding into the interior of the vesicle (Figure \ref{fig:wrapping}d-f).
Co-localization of the membrane fluorescence with the particle fluorescence further corroborates the fully wrapped state. 

\begin{figure*}[ht]  
	\centering{\includegraphics{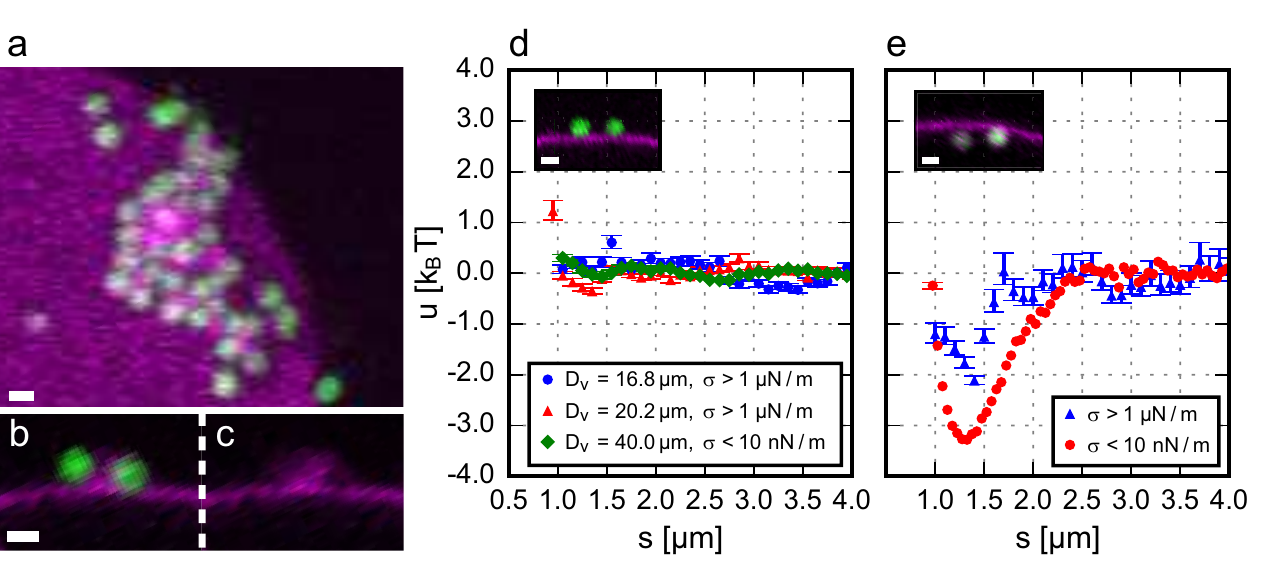}}
	\caption{Interactions between membrane-attached particles. (a)-(c) Confocal images of non-wrapped particles that stick irreversibly together via lipid structures. The membrane fluorescence from the image in (b) is shown separately in (c). (d)-(e) Interaction energy $u(s)$ as a function of geodesic particle separation distance $s$ for (d) two non-wrapped particles and (e) two wrapped particles.
		For non-wrapped particles (d) there is no significant interaction on both tense ($\sigma > \SI{1}{\micro N \per \meter}$) and floppy ($\sigma < \SI{10}{nN \per \meter}$) membranes in a vesicle diameter range of D\textsub{v} = \SIrange{16.8}{40}{\um}.
		For wrapped particles (e) the interaction potential shows a long-ranged attraction.
		The data for tense vesicles is obtained from particle trajectories on a single tense membrane with D\textsub{v} = \SI{36}{\um}, while the interaction energy for floppy vesicles is obtained from an average transition probability matrix of particle trajectories on 3 floppy membranes with D\textsub{v} = \SIrange{14}{40}{\um}.
		Every measurement point is based on \SIrange{20}{1400} independent pair measurements.
		Error bars denote one standard deviation. Scale bars are \SI{1}{\um}.}
	\label{fig:u}
\end{figure*}

This two-state behaviour is in agreement with theoretical predictions \cite{Bahrami2014a}. Using a similar approach, we express the total energy of particle wrapping derived from the Canham-Helfrich energy functional \cite{Helfrich1973}:
\begin{equation}  
E = \bigg(\frac{2\kappa}{R^2} + \sigma - u_{ad} \bigg) A .
\label{eq:E}
\end{equation}
Here, $\kappa$ denotes the membrane bending rigidity, $R$ the particle radius, $\sigma$ the membrane tension, $A$ the contact area, and $u_{ad}$ the adhesion energy per unit area.
This equation states that the energy is minimized by either minimizing or maximizing the contact area, depending only on the sign of the prefactor between brackets.
The value of this prefactor in turn depends on the tunable parameters membrane tension and adhesion energy.

We vary the membrane tension $\sigma$ by adjusting the salt concentration in the vesicle exterior.
For this we discern two extreme situations: \emph{tense} vesicles with a (non-fluctuating) spherical shape ($\sigma > \SI{1}{\micro N \per \meter}$) and \emph{floppy} vesicles that exhibit clear fluctuations around a spherical shape ($\sigma < \SI{10}{nN \per \meter}$).
The values of the surface tension have been derived from the spectral analysis of the fluctuating vesicle contour according to \cite{Pecreaux2004}.

In order to vary the adhesion energy $u_{ad}$, we coat particles with different amounts of linker protein avidin.
We measure the distribution of linker densities in a fluorescence assay (see Method section) and relate this to the fraction of particles that are wrapped by floppy membranes (see Figure \ref{fig:wrapping}g and S1).
We find that wrapping occurs above a critical linker density of \SI{513+-77}{\per\square\um}.
On tense membranes we never observe wrapping of particles.

Our DOPC membranes have membrane bending rigidity $\kappa = \SI{21}{k_BT}$ \cite{Rawicz2000}.
In the case of floppy membranes (\hbox{$\sigma \ll 2\kappa R^{-2}$}) Equation \ref{eq:E} yields a corresponding adhesion energy per unit area of \SI{168}{k_BT\per\square\um}.
Comparing this adhesion energy to the literature value of the binding energy per avidin-biotin bond ($\SI{17}{k_BT}$ \cite{Moy1994}), we conclude that effectively only 2 \% of the surface linkers are binding.
This is probably due to the presence of the polymer between biotin and lipids: the bulky polymer may reduce the binding energy per linker, prevent access to some avidin binding sites, and cause an additional non-specific steric repulsion because of overlap with polymers on the particles.

Note that while wrapping requires floppy membranes, it is irreversible and affects the membrane tension: an initially floppy membrane gradually increases its surface tension upon wrapping particles, due to the effective removal of membrane surface area.
In this way, a tense membrane with wrapped particles can be obtained as well.

The observed ``all or nothing'' wrapping behaviour provides a means to control local membrane deformations through easily accessible experimental parameters.
We will use this experimental control in the next section to investigate the forces between local membrane deformations.

\subsection*{Membrane shape mediated interactions}
When two wrapped particles approach within a distance of several particle diameters, we observe a reversible, long-ranged attraction between them (see Supplementary Video S2).
Excitingly, this interaction is absent for particles that are adhered to but not wrapped by the membrane (see Supplementary Video S3).
This implies that the interaction observed between wrapped particles is purely caused by the local deformation arising from particle wrapping.

To be able to single out the membrane-mediated force, we exclude all other relevant forces on the particles.
Firstly, electrostatic interactions are screened up to a Debye-H\"uckel screening length of \SI{1}{nm} by adding \SI{50}{mM} of salt to the vesicle exterior.
Secondly, gravity is compensated with buoyancy by carefully increasing the water density with heavy water (D\textsub{2}O).
Thirdly, wall interactions are suppressed by coating the coverglass with polymer \cite{Lau2009}.
Finally, we ensure that particles do not attract via Van der Waals forces by grafting a high density of poly(ethylene) glycol (PEG) to the particle surface, which acts as a steric stabilizer \cite{Upadhyayula2012}.
We confirm with particle tracking in three-dimensional confocal images that the particles indeed do not interact or sediment (see Supplementary Fig. S2 online).

In order to quantify the membrane shape mediated interaction, we track the membrane-adhered particles using confocal microscopy at a frame rate of \SIrange{29}{57}{Hz}.
We are able to extract the 3D particle coordinates from these 2D image sequences by simultaneously tracking the vesicle, to which the particles are confined (see Method).
The pair interaction energy is inferred from direct measurement of the transition probability matrix $P_{ij}$, describing the probability for particles to move from separation distance $s_i$ to separation distance $s_j$ \cite{Crocker1994}.
Here, the distance $s$ is the geodesic distance between the points where the particles connect to the membrane.
From $P_{ij}$ a stationary probability distribution for $s$ is obtained.
This is equal to the equilibrium distribution, assuming that the hydrodynamic drag forces on the particle do not depend on their separation.
From the Boltzmann distribution we then determine the energy of two interacting particles, $u(s)$.

Using this method, we infer the pair interaction energy $u(s)$ between wrapped particles and between non-wrapped particles.
Clearly, non-wrapped particles do not interact, while wrapped particles show a long-ranged attraction (Figure~\ref{fig:u}d-e).
The shape of the interaction potential for wrapped particles does not depend on the membrane tension, although the interaction strength is lower on tense membranes.
We find that the interaction strength for floppy membranes is \SI{-3.3}{k\textsub{B}T} and that the attraction extends over a range of \SI{2.5}{\um}, which is equivalent to 2.5 particle diameters.
As the interaction energy is larger than k\textsub{B}T, this attraction can be observed by eye from the relative movement of membrane-wrapped particles in Supplementary Video S2.
The interaction force is only present for particles that deform the membrane: therefore we conclude that the reason for the interaction is the membrane deformation only and that the membrane mediates this force.

In earlier work by Koltover \textit{et al.} \cite{Koltover1999} and Ramos \textit{et al.} \cite{Ramos1999}, an attraction between membrane-bound particles was observed that lead to irreversible aggregation.
Strikingly, in our experiments the interaction potential does not feature this short-ranged permanent binding, but a long-ranged reversible attraction.
We explain this as follows: similar to the referenced work \cite{Koltover1999, Ramos1999}, our initial experiments also contained particle aggregates on the membrane (see Figure \ref{fig:u}a-c and supplementary Fig. S3 online).
However, confocal microscopy revealed that these aggregates are mediated by small ($< \SI{1}{\um}$) lipid vesicles always present in GUV solutions \cite{Tamba2011}.
As their membrane composition is equal to that of the GUVs, they contain biotin linkers as well and thereby irreversibly bind to the adhered particles.
This ``bridging'' process gives rise to a short ranged irreversible attraction.
Previously, these lipid structures could not be identified because they are invisible in bright field microscopy due to their small size, while they are detected easily in the confocal microscopy experiments presented here. 
We deliberately remove the small lipid structures in our experiments by filtration (see Methods section) enabling us to single out the membrane-mediated interaction.

We note again that lipid membranes are profoundly different from liquid/air and liquid/liquid interfaces where surface bending is negligible and surface tension effects such as capillary forces dominate. In the absence of gravity, two ideal spheres bound to a liquid/air or liquid/liquid interface do not interact, because on these types of surfaces, a sphere will adjust its height until it accommodates the wetting angle and therefore does not induce any surface deformation\cite{Binks2006a,Sharifi-mood2015}. Capillary forces due to interface deformations are only observed  for particles with an anisotropy in their shape or roughness of the particle-interface contact line\cite{Binks2006a,Cavallaro2011,Loudet2005,Sharifi-mood2015}. In the case of lipid membranes, however, this consideration is only valid in the limit of high membrane tension, as hypothesized recently by Sarfati and Dufresne\cite{Sarfati2016a}. For our low tension membranes, curvature energy is the most significant energy contribution, as will be corroborated further in the next section.

Analytic approximations for a membrane-bending mediated interaction in the weakly curved limit predict a fluctuation mediated attraction as well as a bending mediated repulsion \cite{Goulian1993a, Dommersnes2002}, at least in the case of isotropic deformations.
Our attractive interaction, however, cannot be caused by fluctuations since such an attraction is negligible compared to k\textsub{B}T at this length scale \cite{Lin2011a}.
The repulsion due to membrane curvature should thus be dominant, but clearly cannot explain the attraction we observe.
Therefore, we conclude that the deformations induced by the wrapped colloidal particles cannot be described by linearised theory.

With non-linear field theory it is possible to calculate the interaction force from the exact membrane shape, even in the highly curved limit \cite{Muller2005}.
Without this information, however, it is not even clear whether the particles repel or attract.
In the limit of asymptotically flat membranes, Reynwar \textit{et al.} \cite{Reynwar2011} computed the interaction energy explicitly by numerically solving the membrane shape equation.
They found an attraction with a well depth of the interaction potential on the same order of magnitude as in our experiments, \SI{-3}{k\textsub{B}T}, albeit with an additional energy barrier at longer ranges.
To more closely resemble our experimental system, we therefore performed computer simulations on spherical membranes with fixed area.

\subsection*{Simulations of membrane mediated interactions}
To investigate the origin of the observed membrane mediated interaction, we simulate the interaction between two particles adhered to a spherical fluid membrane.
Our approach is based on earlier work by {\v{S}}ari{\'{c}} and Cacciuto \cite{Saric2013a,Saric2012,Saric2012a} and is explained in detail in the method section.
In short, we describe the vesicle using a dynamically triangulated network consisting of 5882 vertices.
Between the vertices we apply hard-core repulsion such that the minimum edge length of the network is $l$.
The fluid nature of the membrane is taken into account by allowing the edges of this network to flip.
The vesicle itself, in equilibrium, forms a sphere of diameter $D_{v}=50l$.
We introduce two colloidal particles with diameter $D_{p}=8l$, chosen such that the $D_p/D_v$ ratio is similar to the experimental value.
Having set the volume and surface area of the vesicle to the target values, we apply an adhesion potential between the attractive part of the particles (which in our system is about $90\%$ of the particles' total area) and the vertices in the vesicle to let the membrane wrap around the particles.
We use a Monte Carlo annealing algorithm to identify the equilibrium shape of the membrane for different positions of the particles.
Note that there is no absolute length scale involved in these simulations.

\begin{figure}[ht]  
	\centering{\includegraphics{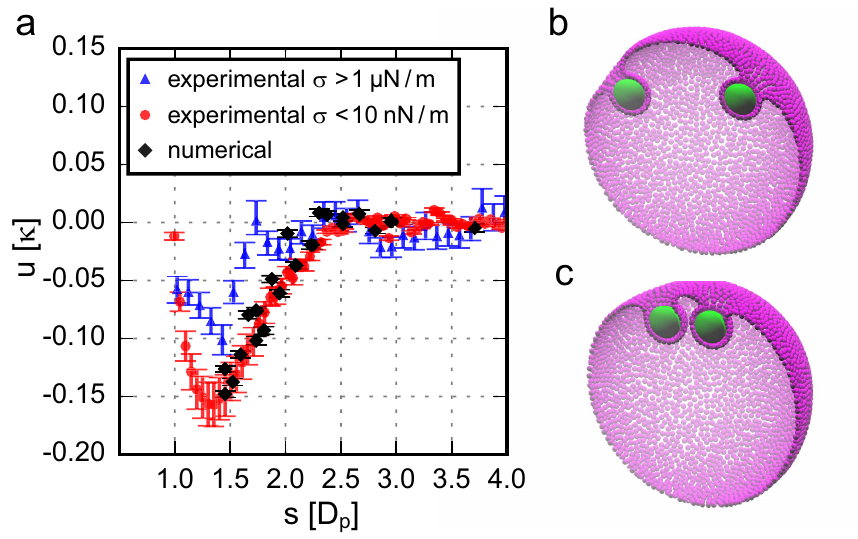}}
	\caption{Computer simulation of the interaction between two membrane-wrapped particles.
		(a) The numerical results are displayed together with the experimental results, which are rescaled by the bending rigidity $\kappa = \SI{20.9+-2.5}{k_BT}$ \cite{Rawicz2000} and the particle diameter $\mathrm{D_p} = \SI{0.98+-0.03}{\um}$.
		(b)-(c) show cross-sections of the simulations of two particles (green) adhered to a coarse-grained membrane (magenta), at separation $s=\SI{3.7}{D_p}$ (b) and $\SI{1.5}{D_p}$ (c).
		The membrane vertices are denoted by small spheres with diameter $l$.}
	\label{fig:simulation}
\end{figure}

As shown in Figure~S4, once the particles are wrapped by the membrane, the adhesion energy and degree of wrapping remain constant, but the curvature energy depends on the separation between the particles.
The excess energy of the membrane for different separations $s$ is shown in Figure~\ref{fig:simulation}, together with the experimental data.
We find that the curvature energy of the membrane favours attraction between the particles for distances $1.5D_p \lesssim s \lesssim 2.2D_p$, in excellent agreement with the experimental results.
For larger distances the energy of the vesicle is barely affected by a change of the separation between the particles.
The minimum distance is set by the resolution of our coarse grained description of the membrane: at $1.5 D_p$ we can be sure to always have two layers of vertices between the particles.
Because of this limitation our simulations cannot capture the short-range effects observed in the experiment as well as in the higher resolution simulations of Reynwar \textit{et al.}~\cite{Reynwar2007,Reynwar2011}.
In contrast to earlier work, however, our simulations do take the overall curved shape of the vesicle into account, as well as the fact that it is a closed surface with fixed area and enclosed volume.
For these conditions, we find that there is no long-range repulsion between the particles, in contrast to some earlier numerical predictions on asymptotically flat membranes~\cite{Reynwar2011}.
The observed attraction is entirely due to a decrease in the bending energy of the membrane upon approach of the particles and  is quantitatively agreeing with our experimental results on vesicles.

\section*{Conclusion}
We established an experimental system of particles adhering to Giant Unilamellar Vesicles that enables quantitative measurement of interactions mediated by lipid membrane curvature. 
For individual membrane-adhering particles, we showed that there are only two states of membrane deformation: a non-wrapped and a fully wrapped state. 
This ``all-or-nothing'' behaviour is controlled by the particle-membrane adhesion strength and membrane tension, which agrees with a simple model based on bending, tension and adhesion energies.

The two-state particle wrapping allows us to selectively measure the effect of local membrane deformations on the pair interaction. 
For two membrane-wrapped particles, we observed a reversible attraction of three times the thermal energy over a distance of several microns.
As the interaction is absent for non-wrapped particles, we conclude that it is mediated by the lipid membrane and originates solely from the particle-induced membrane deformation.

To further probe the underlying physical cause of this attraction, we used a coarse-grained numerical model and Monte Carlo method from which we determined the interaction energy between two wrapped particles.
Apart from the geometry and bending modulus of the membrane, which we respectively set to the experimental and literature values, the model requires no adjustable parameters.
The energy profile we obtained from our simulations is in excellent quantitative agreement with the experiments.
In particular, we find a long-ranged attraction between the wrapped particles, which is entirely due to a decrease in the bending energy of the membrane as the two particles approach each other.

As the observed interaction is determined by bending energy only, there is no absolute length scale involved in the simulations. This implies that the membrane curvature mediated force equally applies to the described colloidal particles as well as to membrane proteins such as proteins containing a BAR domain\cite{Peter2004a}. In fact, our experimental measurements quantitatively model protein interactions on closed membranes, as long as the Helfrich energy description holds. More local membrane deformations induced by for instance transmembrane proteins\cite{McMahon2005} might also deform the membrane in a similar fashion, but on a length scale comparable to the membrane thickness. In this case the membrane cannot be described anymore by a two-dimensional surface, and consequently other effects such as membrane thickness modulations\cite{dan1993} could lead to interactions. On the other hand, aggregates of transmembrane proteins may again act as larger membrane-deforming objects that are described by our model.

To more closely mimic biological systems, lipid membranes with multiple components and colloidal particles with anisotropic shapes or adhesion patches may be employed in the future.  Multi-component membranes may locally phase-separate and affect the measured interaction potential through an intricate process: the object-induced curvature may influence the local membrane composition\cite{Callan-Jones2011} and thus spatially modulate the elastic constants of the membrane. A preference of the particle for a particular phase may further induce a local phase-separation that adds an additional force driven by the line tension of the phase boundaries\cite{Katira2015}. Besides multi-component membranes, colloidal particles with anisotropic shapes or site-specific adhesion patches may be used to mimic the complicated deformation profiles of proteins.  These additional complexities will enable quantitative modeling of the interaction profile between membrane proteins of various geometries and thus further improve our understanding of cellular processes that involve membrane-shaping proteins.

\section*{Methods}
\subsection*{Chemicals}
Styrene, itaconic acid, 4,4'-Azobis(4-cyanovaleric acid) (ACVA),
1,3,5,7-tetramethyl-8-phenyl-4,4-difluoro\-bora\-diaza\-indacene (BODIPY),
methoxypoly(ethylene) glycol amine (mPEG-NH\textsub{2}, M\textsub{w} = 5000),
N-hydroxysulfosuccinimide sodium salt (Sulfo-NHS),
sodium phosphate,
D-glucose,
methanol,
ethanol,
acetic acid,
ammonium hydroxide 28-30\%w/w (NH\textsub{4}OH),
Hellmanex III,
Pluronic F-127,
deuterium oxide 70\% (D\textsub{2}O),
3-(trimethoxysilyl)propyl methacrylate (TPM), and
Bovine Serum Albumin (BSA)
were purchased from Sigma-Aldrich; 
sodium chloride,
sodium azide,
hydrogen peroxide 35\%w/w (H\textsub{2}O\textsub{2}),
acrylamide,
N,N,N',N'-tetramethylethylenediamine (TEMED), and
ammonium persulfate (APS)
from Acros Organics;
1-ethyl-3-(3-dimethylaminopropyl) carbodiimide (EDC) from Carl Roth;
NeutrAvidin from Thermo Scientific;
DNA oligonucleotides (biotin-5'-TTTAATATTA-3'-Cy3) from Integrated DNA Technologies;
$\Delta$9-cis 1,2-dioleoyl-sn-glycero-3-phosphocholine (DOPC),
1,2-dioleoyl-sn-glycero-3-phosphoethanolamine-N-(lissaminerhodamine B sulfonyl) (DOPE-rhodamine), and
1,2-dioleoyl-sn-glycero-3-phospho\-ethanolamine-N-[biotinyl(polyethylene glycol)-2000] (DOPE-PEG-biotin)
from Avanti Polar Lipids.
Unless stated otherwise, chemicals were used as received.
Deionized water is used with \SI{18.2}{M\ohm cm} resistivity, obtained using a Millipore Filtration System (Milli-Q Gradient A10).

\subsection*{Probe particles}
Polystyrene particles are synthesized from styrene, itaconic acid, ACVA, and BODIPY in water using a surfactant-free radical polymerization described in \cite{Appel}, resulting in monodisperse spheres with a diameter of \SI{0.98+-0.03}{\um} (see supplementary Fig. S4 online for a scanning electron microscopy image).
Resulting particles are coated with NeutrAvidin and mPEG-NH\textsub{2} using an a protocol adjusted from \cite{Meng2004}.
All subsequent reactions are done at \SI{4}{\degreeCelsius} to slow down NHS hydrolysis.
\SI{1}{mL} 20\%w/w particles are mixed with \SI{80}{\umol} EDC and \SI{25}{\umol} Sulfo-NHS in \SI{10}{mL} water at pH = 5.3 and stirred for \SI{30}{\minute}.
The pH of the resulting NHS-activated particles is brought to 8.6 using \SI{0.2}{M} NaOH.
\SI{750}{\uL} of the 2\%w/w activated particles is then mixed with \SIrange{0.5}{50}{\ug} NeutrAvidin.
After \SI{30}{\minute}, \SI{4}{mg} mPEG-NH\textsub{2} is added and the reaction proceeds for \SI{40}{\hour}.
Then the pH is brought to 12 with \SI{1}{M} NaOH, the particles are ultrasonicated for \SI{5}{\minute} and then washed 1 time with \SI{0.01}{M} HCl and 3 times with water.
Finally, sodium azide is added to a concentration of \SI{3}{mM} to prevent bacterial growth.

\subsection*{Biotin binding sites assay} In order to quantify the number of biotin binding sites (the `linker density') on each particle, we measure fluorescence of biotin- and dye-functionalised DNA strands.
DNA strands have the advantage that they are well soluble in water, so that there is no non-specific adhesion to the particle surface.
\SI{10}{\uL} \SI{6}{\uM} DNA (in water) is mixed with \SI{10}{\uL} 0.5\%w/w particles in a total volume of \SI{310}{\uL} \SI{50}{mM} PBS buffer with 0.5\%w/w Pluronic F-127.
The mixture is heated to \SI{55}{\degreeCelsius} for \SI{30}{\minute} and washed 3 times with water.
The sample is diluted 10 times in a PBS buffer inserted into an untreated rectangular glass capillary, which immobilizes the particles.
The fluorescence intensity is quantified using fluorescence microscopy (Nikon Intensilight) with reproducible settings and using a reference value obtained from commercial particles with a known amount of avidin (Spherotech PC-S-1.0) we can obtain a distribution of avidin linkers per particle.

\subsection*{GUV preparation}
Vesicles are prepared using a standard electroswelling technique \cite{Angelova1986a}.
A lipid mixture of 97.5\%w/w DOPC, 2\%w/w DOPE-PEG-biotin, and 0.5\%w/w DOPE-rhodamine is used, ensuring a liquid bilayer at room temperature.
2x\SI{20}{\ug} of the lipids in chloroform are dried on two \SIlist[list-pair-separator=x]{25;25}{mm} ITO-coated glass slides (\SIrange{15}{25}{\ohm}, Sigma-Aldrich), placed in \SI{1.8}{mL} of a solution with \SI{100}{mM} glucose and \SI{0.3}{mM} sodium azide in 49:51 (mass) D\textsub{2}O:H\textsub{2}O.
The cell is subjected to \SI{1.1}{V} (rms) at \SI{10}{Hz} for \SI{2}{\hour}, with the first \SI{2}{\minute} a linear increase from \SI{0}{V}. GUVs are stored in a BSA-coated glass vial at room temperature.
In order to remove small lipid structures \cite{Tamba2011}, \SI{100}{\uL} GUV solution is pipetted on a Whatmann \SI{5.0}{\um} pore size cellulosenitrate filter and slowly flushed with \SI{5.0}{\mL} of glucose solution.
\SI{100}{\uL} purified GUVs are harvested from the filter and used the same day.
All handling is done with care not to mechanically shock the solution.

\subsection*{Coverglass treatment}
We employed a polymerization of acrylamide onto TPM-coated glasses \cite{Lau2009}, as follows: coverglasses are cleaned for \SI{30}{\minute} in a 2\%v/v Hellmanex solution, rinsed 3 times with water, immersed in 5:1:1 H\textsub{2}O:NH\textsub{4}OH:H\textsub{2}O\textsub{2} for \SI{30}{\minute} at \SI{70}{\degreeCelsius}, rinsed 3 times with water, and 2 times with ethanol.
TPM functionalisation is done by immersing \SI{15}{\minute} in ethanol with 1\%v/v acetic acid and 0.5\%v/v TPM, rinsing 3 times with ethanol and incubating for \SI{1}{\hour} at \SI{80}{\degreeCelsius}. Polymerization is done in a 2\%w/w solution of acrylamide (evacuated in vacuum for \SI{30}{\minute} to remove oxygen), with 0.035\%v/w TEMED and 0.070\%w/w APS for \SI{2}{\hour}. Resulting coverglasses were kept inside the polymerization solution at \SI{4}{\degreeCelsius} until use. Directly before use, a coverglass is rinsed with water and blow-dried with nitrogen. 

\subsection*{Density matching}
A density-matched PBS stock buffer of \SI{200}{mOsm} is prepared containing \SI{10.0}{mM} sodium phosphate, \SI{82.0}{mM} sodium chloride, and \SI{3.0}{mM} sodium azide.
Density-matching with the probe particles was achieved by gradually adding D\textsub{2}O until no sedimentation or creaming occurred at \SI{10000}{g} for \SI{1}{\hour}.
The mass ratio D\textsub{2}O:H\textsub{2}O for water is roughly 51:49; for the buffer it is roughly 45:55 because the solutes increase the density.
Using ratios between the stock buffer and density-matched water, buffers at different osmolarities are obtained.
Density matching is confirmed for each mixture separately.

\subsection*{Sample preparation}
Samples are prepared on polyacrylamide-coated coverglasses and density matched with D\textsub{2}O.
\SI{2}{\uL} 2\%w/w particles, \SI{4}{\uL} \SI{150}{mOsm} PBS buffer, and \SI{20}{\uL} filtered GUVs are incubated for \SI{10}{\minute} in a plastic microtube.
Then \SI{10}{\uL} of this mixture is slowly distributed into the sample holder with \SI{50}{\uL} \SI{100}{mOsm} PBS buffer already inside.
The sample holder consists of a Teflon ring clamped on a pretreated coverglass.
For tense GUVs, the sample holder is closed with vacuum grease and a second coverglass; for floppy GUVs, the sample holder is kept open to air for \SI{30}{\minute} so that evaporation leads to an increase in osmotic pressure, and consequentially a decrease of membrane tension.
All experiments were performed at a room temperature of \SIrange{19}{22}{\degreeCelsius}.

\subsection*{Imaging}
Imaging is done with an inverted Nikon TiE microscope equipped with a Nikon A1R confocal scanhead with both galvano and resonant scanning mirrors.
High-speed trajectory imaging is achieved with a horizontal resonant mirror scanning lines at \SI{15}{kHz}; single-particle close-ups are done with the galvano mirrors.
We use a 60x water immersion objective (NA = 1.2) to reduce axial aberration due to index of refraction mismatch.
The excitation laser is passed through a quarter wave plate to mitigate polarization effects of bilayer-attached dye molecules.
Excitation (at \SI{488}{nm} and \SI{561}{nm}) and detection are performed simultaneously (for trajectory imaging) or sequentially (for close-ups) using a dichroic mirror splitting the emission signal onto \SIrange{500}{550}{nm} and \SIrange{565}{625}{nm} filters.
The sample is mounted on an MCL NanoDrive stage to enable fast Z stack acquisition.

\subsection*{Image analysis}
The raw images of the interaction measurements are high speed (\SIrange{29}{57}{Hz}) confocal images containing two separate colours, being the vesicle fluorescence and the particle fluorescence.
Images are convolved with a Gaussian kernel with an rms width of 1 pixel to reduce random noise.

The particle fluorescence signals are tracked using a widely employed centre-of-mass based particle tracking algorithm in a Python implementation \cite{Crocker1996}, which is available online \cite{Allan2015}.
All particle trajectories are checked manually for missing coordinates and corrected if necessary.
Because the used centre-of-mass refinement technique systematically finds coordinates of overlapping features too close together, we refine overlapping signals additionally by least-squares fitting to a sum of Gaussians.

The three-dimensional coordinates of the particles relative to the vesicle ($x_{rel}$, $y_{rel}$, $z_{rel}$) are determined as follows: $x_{rel} = x - x_c$; $y_{rel} = y - y_c$; $z_{rel}^2 = (R + h)^2 - x_{rel}^2 - y_{rel}^2$.
The vesicle radius $R$ is obtained in a separate three-dimensional confocal measurement; the particle-vesicle distance $h$ is known from the wrapping state of the particle; the vesicle centre ($x_c$, $y_c$) is measured simultaneously with the particle tracking ($x$, $y$) from the high-speed 2D confocal images.

For the vesicle tracking, we interpolate the image on lines that are drawn outwards from a rough estimate of the vesicle centre.
The maximum value on each of these lines provides an estimate of the vesicle perimeter.
Around each maximum, a fit region of 5 pixels is defined for further refinement: linear regression on the discrete derivative provides the position of the vesicle perimeter with sub-pixel resolution.
Finally, we perform a least-squares fit to a circle (for two dimensions) or ellipsoid (for three dimensions) to obtain the refined vesicle centre and radius.
The here described algorithm is available online \cite{VanderWel2016}.

\subsection*{Modelling Details}
The curvature energy of a biological membrane is described by the Helfrich energy functional\cite{Helfrich1973} as:

\begin{align}
\label{CHenergy}
u_{\mathrm{Curv}}= \frac{\kappa}{2}\int_{A} (2H)^2 dA,
\end{align}
where $H$ is the mean curvature of the membrane, which is defined as the divergence of the surface normal vector, $H= -\frac{1}{2} \nabla \cdot \mathbf{n}$. We model the vesicle by a network of vertices with the minimum length of $l$ that are connected in a triangular network. The curvature energy of our discretised membrane is given by:

\begin{align}
\label{uCurv}
u_\mathrm{Curv} =  \sqrt{3}\kappa \sum_{<ij>}^{} 1- \textbf{n}_i \cdot \textbf{n}_j,
\end{align}
where $\textbf{n}_i$ and $\textbf{n}_j$ are the normal vectors to any pair of adjacent triangles $i$ and $j$, respectively. The summation runs over all pairs of such triangles. To simulate the fluidity of the membrane, we change the connectivity of the network: we cut and reattach connections between the four vertices of any two neighbouring triangles. The surface area $A$ and volume $V$ of the vesicle, are maintained by constraints $u_A= K_A (A-A_t)^2/A_t$ and $u_V= K_V (V-V_t)^2/V_t$ with $K_A = 10^3 k_B T/l^2$ and $K_V =4\times 10^3 k_B T/D_p l^2 $, where $k_BT$, $D_p$, $A_t$ and $V_t$ are the thermal energy, the diameter of the particles, the target surface area and the target volume of the vesicle, respectively. In each simulation we set the target values of surface area and volume of the vesicle with diameter $D_v = 50l$ as $A_t= 1.05 A_0$ and $V_t=V_0$, respectively. These parameters cause the final volume and surface area to deviate less than $0.01 \% $ from the target values. To let the vertices of the membrane wrap around the particles, we introduce an attraction potential between them:

\begin{align}
\label{uAd}
u_{\mathrm{Ad}} = \begin{cases} 
-\varepsilon (l_m /r)^6 & \mbox{if } \theta \leq \theta_\mathrm{Wr}, \\ 0 & \mbox{ otherwise}, \end{cases}
\end{align}
where $\varepsilon$ is the particles' adhesion energy and $r$ is the centre to centre distance between particles and vertices. $\theta$ is the angle between the vector normal to the active area of the particles and the vectors that connect the particles to vertices (see supplementary Fig. S5 online). The maximum angle $\theta_\mathrm{Wr}$ is defined to control the area that is forced to be wrapped by the membrane, preventing very sharp membrane bends. $l_m = (l + D_{p})/2$ is the shortest distance between particles and vertices, where the diameter of the particles is set to $D_{p} = 8 l$. We set a cut-off radius for the attraction potential at $1.2 l$ to make sure that other than forming a layer of membrane on the surface of the particles, it has no extra effects. The total energy $u_\mathrm{T}$ of the system is the sum of the curvature energy (Eq.~\ref{uCurv}) and the adhesion energy (Eq.~\ref{uAd}).

To analyse the equilibrium shape of the membrane, we implement the Monte Carlo simulated annealing in order to minimize the total energy of the system. For our Monte Carlo simulations, we use the Metropolis algorithm to move vertices and particles, and flip the edges of the membrane triangulation, in order to change the configuration of the system (shape of the membrane). The temperature of the system is also slowly decreased so that we suppress the fluctuation of the membrane and identify the minimum-energy configuration.

\section*{Acknowledgements}
This work was supported by the Netherlands Organisation for Scientific Research (NWO/OCW), as part of the Frontiers of Nanoscience program and VENI grant 680-47-431.
We thank Jeroen Appel and Wim Pomp for advice on the protocol design and Marcel Winter and Ruben Verweij for experimental support.

\section*{Author contributions statement}
D.J.K., D.H., and T.I. designed the project; C.v.d.W. carried out experiments and data analysis supervised by D.J.K. and D.H..; A.V. performed simulations supervised by T.I. and A.\v{S}.; all authors contributed to writing and revising the manuscript.

\section*{Additional information}
\textbf{Competing financial interests} The authors declare no competing financial interests.

\newpage
\onecolumn

\setcounter{figure}{0}
\setcounter{table}{0}
\makeatletter
\renewcommand{\fnum@figure}{Figure S\thefigure}
\renewcommand{\fnum@table}{Table S\thetable}
\renewcommand{\theequation}{S\arabic{equation}}
\makeatother

\section*{Supplementary material for ``Lipid membrane-mediated attraction between curvature inducing objects''}

\large Author list: Casper van der Wel, Afshin Vahid, An\dj ela~\v{S}ari\'c, Timon Idema, Doris Heinrich, and Daniela J. Kraft

\section*{Supplementary Videos}
\setcounter{figure}{0}
\makeatletter
\renewcommand{\fnum@figure}{Supplementary Video S\thefigure}
\makeatother

Videos are separately available online. Here, still images of the videos are shown together with their captions.
\begin{figure}[H]
	\centering{\includegraphics[width=0.8\textwidth]{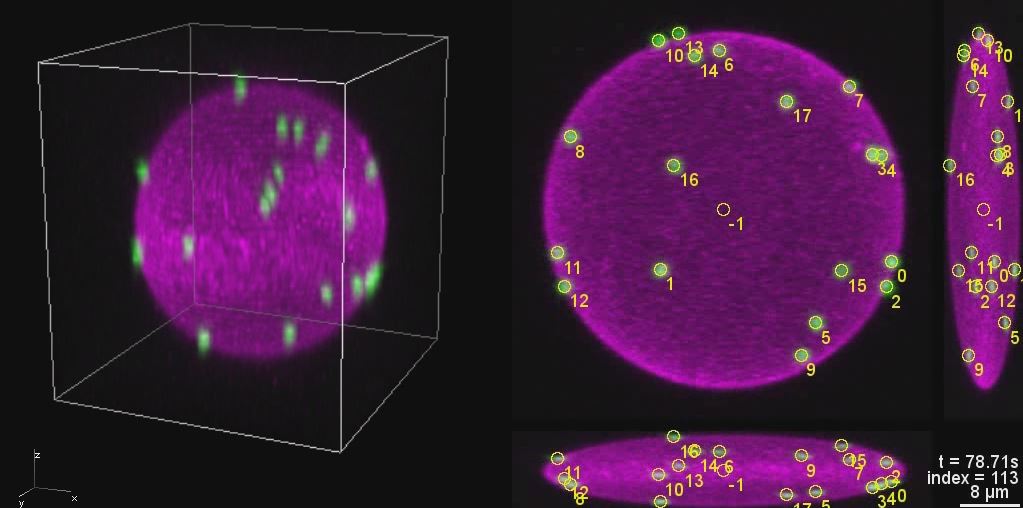}}
	\caption
	{Three-dimensional confocal image sequence of a Giant Unilamellar Vesicle with attached colloidal particles. GUVs are shown in magenta and particles in green. On the left, a three-dimensional rendering is shown. On the right, the particle tracking is shown in an overlay on three maximum intensity projections (centre: xy, bottom: xz, right: yz). The scale bar denotes the pixel size of the xy projection. Z axes are compressed because the physical size of one voxel is larger in the z-dimension than in the xy dimensions. Label ``-1'' denotes the vesicle centre. Particles are \SI{1}{\um} in diameter.}
\end{figure}

\begin{figure}[H]
	\centering{\includegraphics[width=0.8\textwidth]{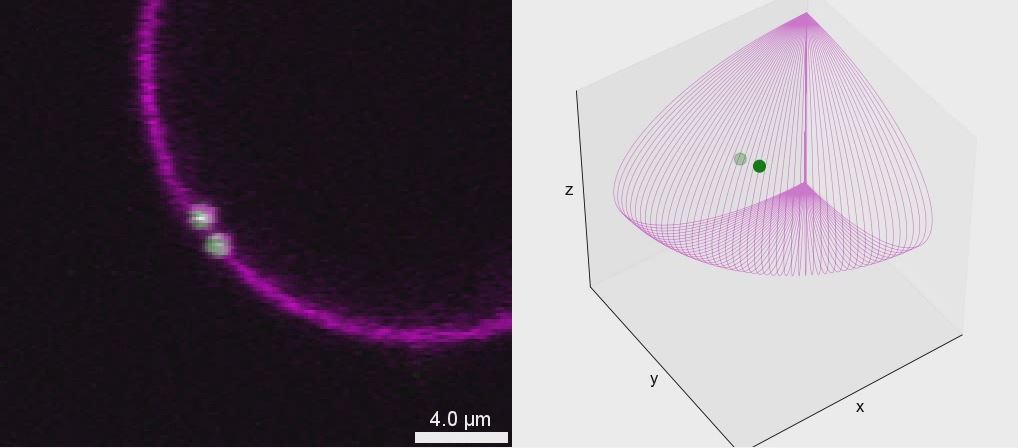}}
	\caption
	{Three wrapped particles on a tense vesicle. From the particle trajectories, it is clear that the particles attract each other. The video shows an image sequence of confocal slices of a spherical vesicle (in magenta) with colloidal particles (in green), displayed in real time. By using information from the particle-vesicle distance and the vesicle radius, the full three-dimensional coordinates of the particles can be reconstructed. This is shown on the right in a three-dimensional rendering. Particles are \SI{1}{\um} in diameter.}
\end{figure}

\begin{figure}[H]
	\centering{\includegraphics[width=0.5\textwidth]{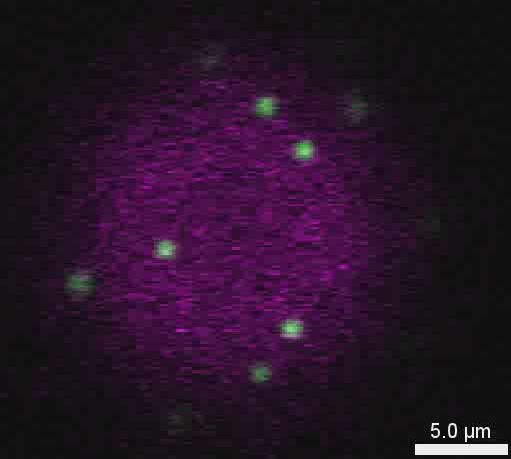}}
	\caption
	{Non-wrapped particles adhered to a vesicle. The particles do not interact with each other. The video shows a confocal image sequence of the top part of a spherical vesicle (in magenta) with colloidal particles (in green), displayed in real time.}
\end{figure}

\section*{Supplementary Figures}
\setcounter{figure}{0}
\makeatletter
\renewcommand{\fnum@figure}{Supplementary Figure S\thefigure}
\makeatother

\begin{figure}[H]
	\centering{\includegraphics{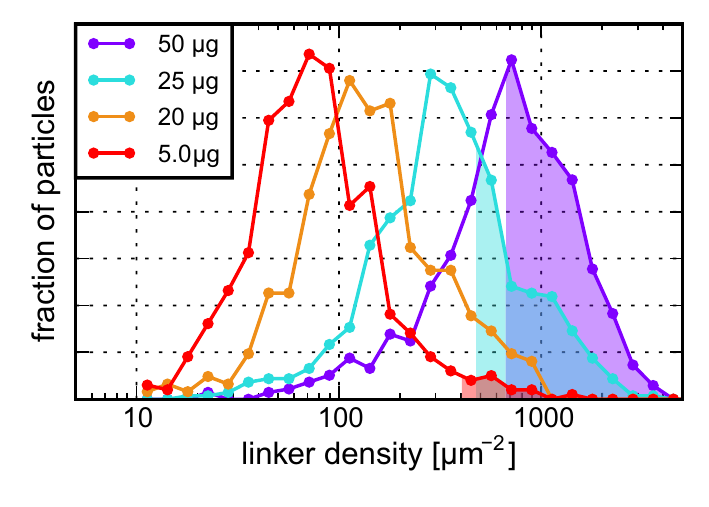}}
	\caption
	{Distribution of particle linker densities for four samples with different amounts of linker protein avidin. The linker amount noted in the legend is the amount added to \SI{15}{mg} particles during synthesis, see Method section. In order to relate the fraction of wrapped particles to these distributions, the right tail of each distribution is shaded up to the measured fraction of wrapped particles. From this, we estimate the critical linker density to be \SI{513+-77}{\per\square\um}. Note that we observed no wrapping for the \SI{20}{\micro \gram} sample (yellow).}
\end{figure}

\begin{figure}[H]
	\centering{\includegraphics{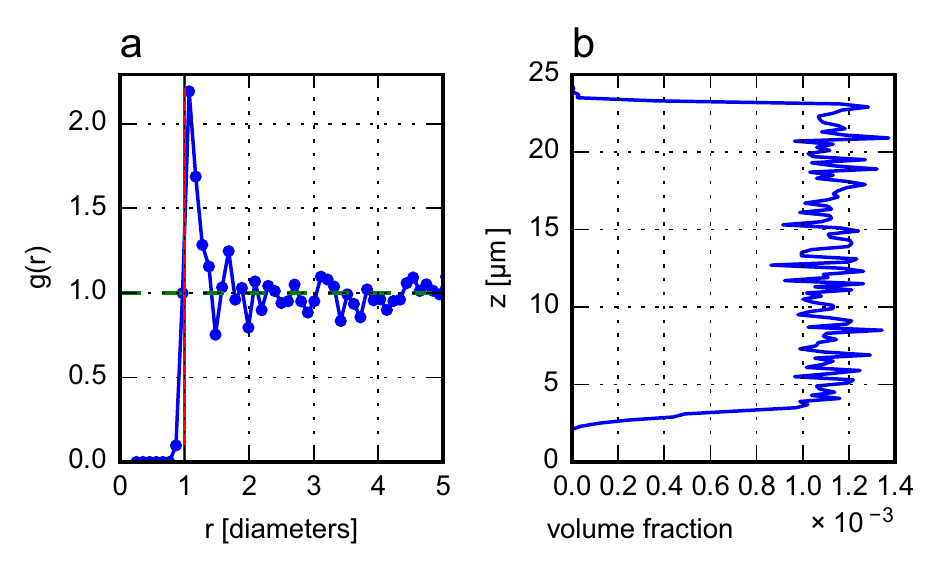}}
	\caption
	{Three-dimensional radial distribution function and sedimentation profile of particles suspended at a volume fraction of 0.0011 in a \SI{50}{mM} density matched PBS solution. (a) The radial distribution g(r) shows no interaction between particles. The red line indicates particle contact. The sharp peak at a distance of 1 diameter is due to the presence of a few dimers originating from the particle synthesis. (b) The density profile shows that there is no gradient in concentration due to gravity.}
\end{figure}

\begin{figure}[H]
	\centering{\includegraphics{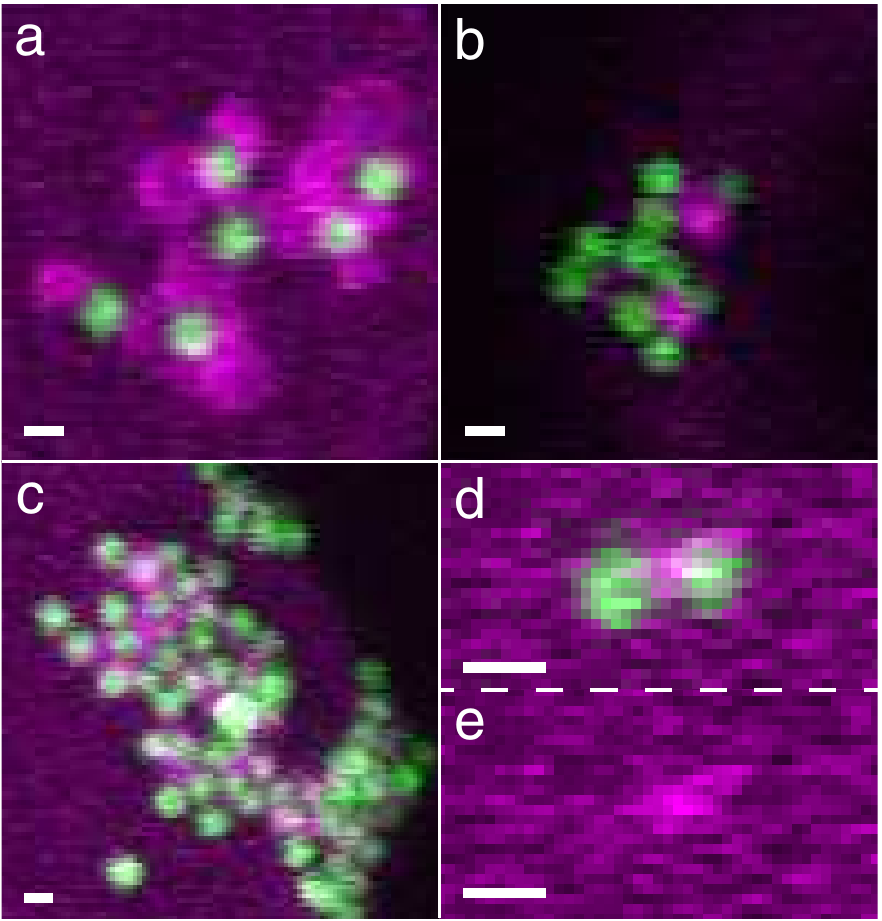}}
	\caption
	{Particle aggregates mediated by small lipid structures. In (a)-(c), permanent particle aggregates are shown (in green) that are mediated by lipid structures, that are visible by their fluorescence (in magenta). In (d)-(e) different fluorescence channels from the same permanent dimer is shown, the bright spot of membrane fluorescence in (e) is the lipid structure that causes the irreversible binding. Scalebars are \SI{1}{\um}.}
\end{figure}

\begin{figure}[H]
	\centering{\includegraphics[width=0.7\textwidth]{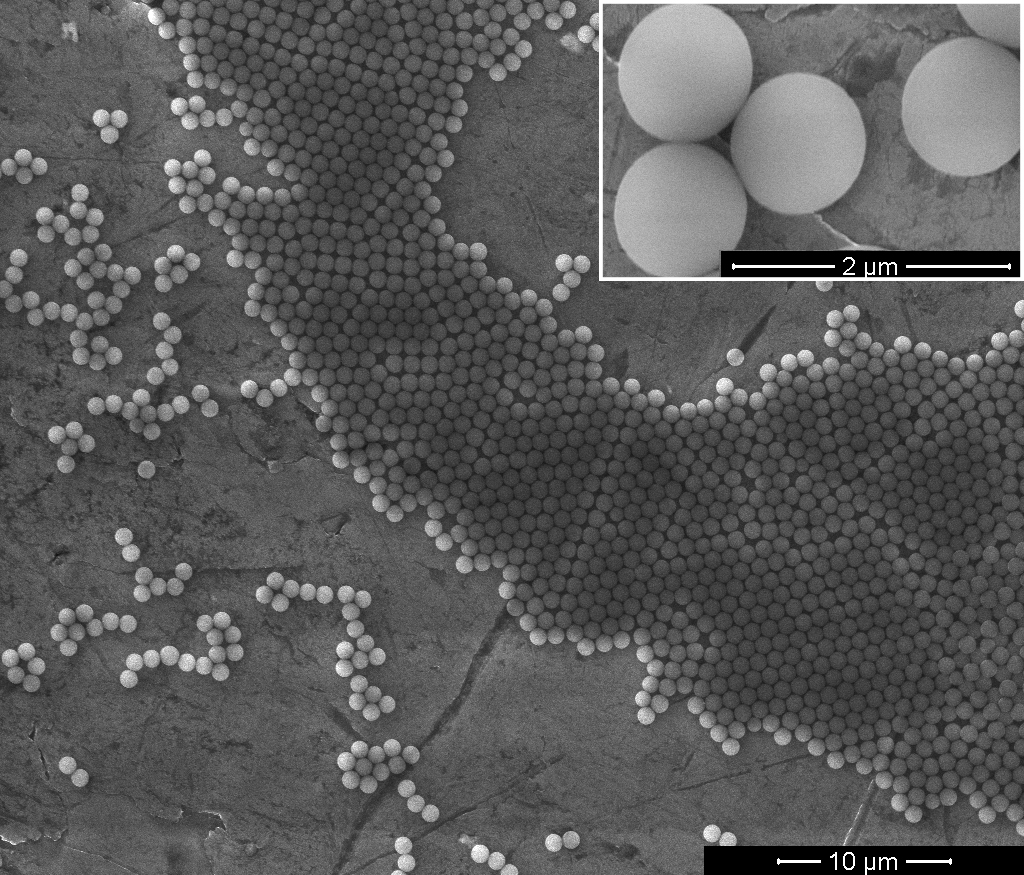}}
	\caption
	{Scanning Electron Microscopy image of the \SI{0.98}{\um} polystyrene colloidal particles used in this work. Images are obtained with an FEI nanoSEM 200 at \SI{15}{kV}. From the two-dimensional crystallization, it is clear that the size polydispersity is low (\SI{0.03}{\um}). The inset shows the smooth surface of the particles.}
\end{figure}

\begin{figure}[H]
	\centering{\includegraphics[width=0.75\textwidth]{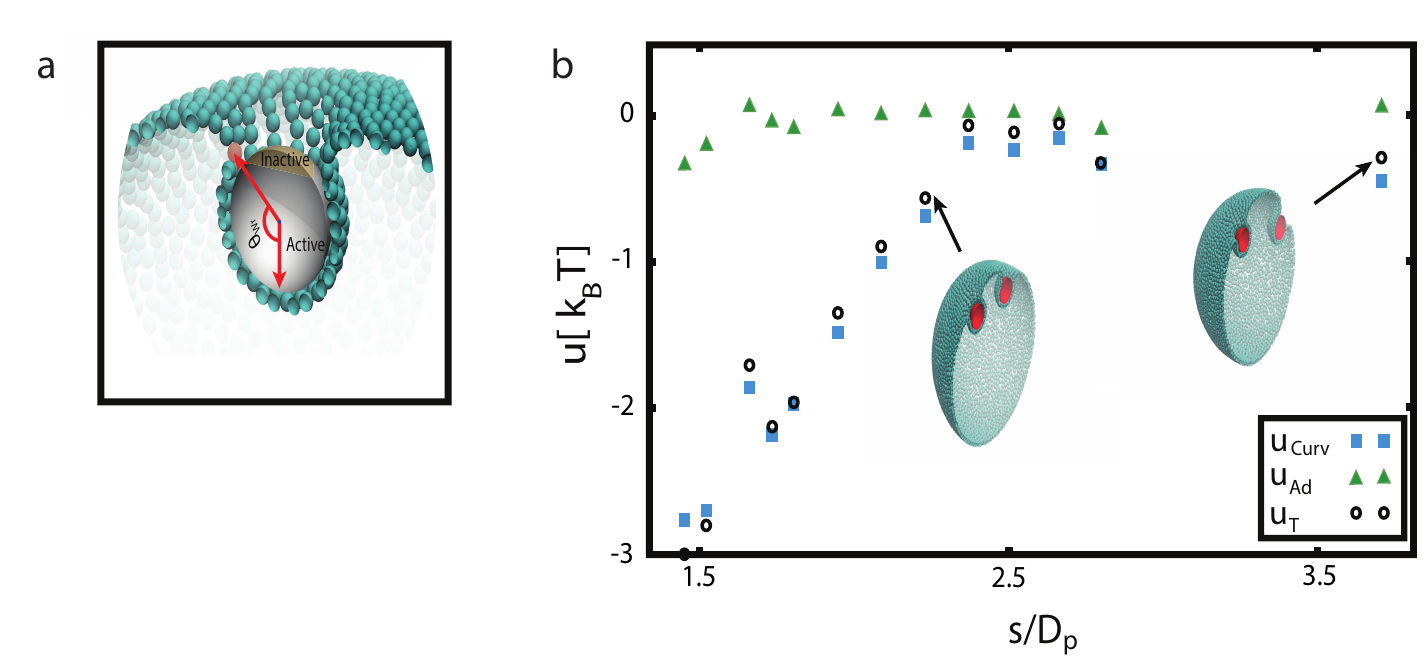}}
	\caption
	{Wrapping of particles by the membrane and the resulting total membrane energy in our numerical model. (a) Wrapping happens through adhesion of membrane vertices to colloid particles, due to a strong adhesion potential (Equation 4). We can specify an inactive region at the top of the colloid, preventing the membrane from making very sharp turns (with very high bending energies); in the given example, $\theta_\mathrm{Wr} = 11\pi/12$. (b) Curvature, adhesion, and total energy of the system, with zero set at the value of two wrapped particles located at opposite poles of the vesicle. After the wrapping process, the adhesion energy ($u_{\mathrm{Ad}}$) does not change significantly and therefore the curvature energy ($u_{\mathrm{Curv}}$) determines the behaviour of the particles.}
	\label{figureS4}
\end{figure}


\begin{thebibliography}{10}
	\expandafter\ifx\csname url\endcsname\relax
	\def\url#1{\texttt{#1}}\fi
	\expandafter\ifx\csname urlprefix\endcsname\relax\def\urlprefix{URL }\fi
	\providecommand{\bibinfo}[2]{#2}
	\providecommand{\eprint}[2][]{\url{#2}}
	
	\bibitem{McMahon2005}
	\bibinfo{author}{McMahon, H.~T.} \& \bibinfo{author}{Gallop, J.~L.}
	\newblock \bibinfo{title}{{Membrane curvature and mechanisms of dynamic cell
			membrane remodelling}}.
	\newblock \emph{\bibinfo{journal}{Nature}} \textbf{\bibinfo{volume}{438}},
	\bibinfo{pages}{590--596} (\bibinfo{year}{2005}).
	
	\bibitem{Freche2011}
	\bibinfo{author}{Freche, D.}, \bibinfo{author}{Pannasch, U.},
	\bibinfo{author}{Rouach, N.} \& \bibinfo{author}{Holcman, D.}
	\newblock \bibinfo{title}{{Synapse geometry and receptor dynamics modulate
			synaptic strength}}.
	\newblock \emph{\bibinfo{journal}{PLoS ONE}} \textbf{\bibinfo{volume}{6}},
	\bibinfo{pages}{e25122} (\bibinfo{year}{2011}).
	
	\bibitem{Pfefferkorn2012}
	\bibinfo{author}{Pfefferkorn, C.~M.}, \bibinfo{author}{Jiang, Z.} \&
	\bibinfo{author}{Lee, J.~C.}
	\newblock \bibinfo{title}{{Biophysics of alpha-synuclein membrane
			interactions}}.
	\newblock \emph{\bibinfo{journal}{Biochim. Biophys. Acta}}
	\textbf{\bibinfo{volume}{1818}}, \bibinfo{pages}{162--171}
	(\bibinfo{year}{2012}).
	
	\bibitem{Goulian1993a}
	\bibinfo{author}{Goulian, M.}, \bibinfo{author}{Bruinsma, R.} \&
	\bibinfo{author}{Pincus, P.}
	\newblock \bibinfo{title}{{Long-range forces in heterogeneous fluid
			membranes}}.
	\newblock \emph{\bibinfo{journal}{Europhys. Lett.}}
	\textbf{\bibinfo{volume}{22}}, \bibinfo{pages}{145--150}
	(\bibinfo{year}{1993}).
	
	\bibitem{Kim1998}
	\bibinfo{author}{Kim, K.~S.}, \bibinfo{author}{Neu, J.} \&
	\bibinfo{author}{Oster, G.}
	\newblock \bibinfo{title}{{Curvature-mediated interactions between membrane
			proteins}}.
	\newblock \emph{\bibinfo{journal}{Biophysical J.}}
	\textbf{\bibinfo{volume}{75}}, \bibinfo{pages}{2274--2291}
	(\bibinfo{year}{1998}).
	
	\bibitem{Dommersnes2002}
	\bibinfo{author}{Dommersnes, P.~G.} \& \bibinfo{author}{Fournier, J.-B.}
	\newblock \bibinfo{title}{{The many-body problem for anisotropic membrane
			inclusions and the self-assembly of ``saddle'' defects into an ``egg carton''}}.
	\newblock \emph{\bibinfo{journal}{Biophysical J.}}
	\textbf{\bibinfo{volume}{83}}, \bibinfo{pages}{2898--2905}
	(\bibinfo{year}{2002}).
	
	\bibitem{Muller2005}
	\bibinfo{author}{M{\"{u}}ller, M.~M.}, \bibinfo{author}{Deserno, M.} \&
	\bibinfo{author}{Guven, J.}
	\newblock \bibinfo{title}{{Interface-mediated interactions between particles: a
			geometrical approach}}.
	\newblock \emph{\bibinfo{journal}{Phys. Rev. E}}
	\textbf{\bibinfo{volume}{72}}, \bibinfo{pages}{061407}
	(\bibinfo{year}{2005}).
	
	\bibitem{Yolcu2014}
	\bibinfo{author}{Yolcu, C.}, \bibinfo{author}{Haussman, R.~C.} \&
	\bibinfo{author}{Deserno, M.}
	\newblock \bibinfo{title}{{The effective field theory approach towards
			membrane-mediated interactions between particles.}}
	\newblock \emph{\bibinfo{journal}{Adv. Colloid Interfac.}}
	\textbf{\bibinfo{volume}{208}}, \bibinfo{pages}{89--109}
	(\bibinfo{year}{2014}).
	
	\bibitem{Reynwar2007}
	\bibinfo{author}{Reynwar, B.~J.} \emph{et~al.}
	\newblock \bibinfo{title}{{Aggregation and vesiculation of membrane proteins by
			curvature-mediated interactions.}}
	\newblock \emph{\bibinfo{journal}{Nature}} \textbf{\bibinfo{volume}{447}},
	\bibinfo{pages}{461--465} (\bibinfo{year}{2007}).
	
	\bibitem{Saric2013a}
	\bibinfo{author}{{\v{S}}ari{\'{c}}, A.} \& \bibinfo{author}{Cacciuto, A.}
	\newblock \bibinfo{title}{{Self-assembly of nanoparticles adsorbed on fluid and
			elastic membranes}}.
	\newblock \emph{\bibinfo{journal}{Soft Matter}} \textbf{\bibinfo{volume}{9}},
	\bibinfo{pages}{6677--6695} (\bibinfo{year}{2013}).
	
	\bibitem{Simunovic2013}
	\bibinfo{author}{Simunovic, M.}, \bibinfo{author}{Srivastava, A.} \&
	\bibinfo{author}{Voth, G.~A.}
	\newblock \bibinfo{title}{{Linear aggregation of proteins on the membrane as a
			prelude to membrane remodeling}}.
	\newblock \emph{\bibinfo{journal}{P. Natl. Acad. Sci. USA}} \textbf{\bibinfo{volume}{110}},
	\bibinfo{pages}{20396--20401} (\bibinfo{year}{2013}).
	
	\bibitem{Pamies2011}
	\bibinfo{author}{P{\`{a}}mies, J.~C.} \& \bibinfo{author}{Cacciuto, A.}
	\newblock \bibinfo{title}{{Reshaping elastic nanotubes via self-assembly of
			surface-adhesive nanoparticles}}.
	\newblock \emph{\bibinfo{journal}{Phys. Rev. Lett.}}
	\textbf{\bibinfo{volume}{106}}, \bibinfo{pages}{045702}
	(\bibinfo{year}{2011}).
	
	\bibitem{Saric2012a}
	\bibinfo{author}{{\v{S}}ari{\'{c}}, A.} \& \bibinfo{author}{Cacciuto, A.}
	\newblock \bibinfo{title}{{Fluid membranes can drive linear aggregation of
			adsorbed spherical nanoparticles}}.
	\newblock \emph{\bibinfo{journal}{Phys. Rev. Lett.}}
	\textbf{\bibinfo{volume}{108}}, \bibinfo{pages}{118101}
	(\bibinfo{year}{2012}).
	
	\bibitem{Vahid2015}
	\bibinfo{author}{Vahid, A.} \& \bibinfo{author}{Idema, T.}
	\newblock \bibinfo{title}{{Point-like inclusion interactions in tubular
			membranes}}.
	\newblock \emph{\bibinfo{journal}{arXiv}} \bibinfo{pages}{1510.03610}
	(\bibinfo{year}{2015}).
	
	\bibitem{Peter2004a}
	\bibinfo{author}{Peter, B.~J.} \emph{et~al.}
	\newblock \bibinfo{title}{{BAR domains as sensors of membrane curvature: the
			amphiphysin BAR structure}}.
	\newblock \emph{\bibinfo{journal}{Science}} \textbf{\bibinfo{volume}{303}},
	\bibinfo{pages}{495--499} (\bibinfo{year}{2004}).
	
	\bibitem{Semrau2009}
	\bibinfo{author}{Semrau, S.}, \bibinfo{author}{Idema, T.},
	\bibinfo{author}{Schmidt, T.} \& \bibinfo{author}{Storm, C.}
	\newblock \bibinfo{title}{{Membrane-mediated interactions measured using
			membrane domains}}.
	\newblock \emph{\bibinfo{journal}{Biophysical J.}}
	\textbf{\bibinfo{volume}{96}}, \bibinfo{pages}{4906--4915}
	(\bibinfo{year}{2009}).

	\bibitem{Koltover1999}
	\bibinfo{author}{Koltover, I.}, \bibinfo{author}{R{\"{a}}dler, J.~O.} \&
	\bibinfo{author}{Safinya, C.~R.}
	\newblock \bibinfo{title}{{Membrane mediated attraction and ordered aggregation
			of colloidal particles bound to giant phospholipid vesicles}}.
	\newblock \emph{\bibinfo{journal}{Phys. Rev. Lett.}}
	\textbf{\bibinfo{volume}{82}}, \bibinfo{pages}{1991--1994}
	(\bibinfo{year}{1999}).

	\bibitem{Ramos1999}
	\bibinfo{author}{Ramos, L.}, \bibinfo{author}{Lubensky, T.},
	\bibinfo{author}{Dan, N.}, \bibinfo{author}{Nelson, P.} \&
	\bibinfo{author}{Weitz, D.}
	\newblock \bibinfo{title}{{Surfactant-mediated two-dimensional crystallization
			of colloidal crystals}}.
	\newblock \emph{\bibinfo{journal}{Science}} \textbf{\bibinfo{volume}{286}},
	\bibinfo{pages}{2325--2328} (\bibinfo{year}{1999}).
	
	\bibitem{Loudet2005}
	\bibinfo{author}{Loudet, J.~C.}, \bibinfo{author}{Alsayed, A.~M.},
	\bibinfo{author}{Zhang, J.} \& \bibinfo{author}{Yodh, A.~G.}
	\newblock \bibinfo{title}{{Capillary interactions between anisotropic colloidal
			particles}}.
	\newblock \emph{\bibinfo{journal}{Phys. Rev. Lett.}}
	\textbf{\bibinfo{volume}{94}}, \bibinfo{pages}{018301}
	(\bibinfo{year}{2005}).

	\bibitem{Cavallaro2011}
	\bibinfo{author}{{Cavallaro, Jr.}, M.}, \bibinfo{author}{Botto, L.},
	\bibinfo{author}{Lewandowski, E.~P.}, \bibinfo{author}{Wang, M.} \&
	\bibinfo{author}{Stebe, K.~J.}
	\newblock \bibinfo{title}{{Curvature-driven capillary migration
			and assembly of rod-like particles}}.
	\newblock \emph{\bibinfo{journal}{P. Natl. Acad. Sci. USA}} \textbf{\bibinfo{volume}{108}},
	\bibinfo{pages}{20923--20928} (\bibinfo{year}{2011}).
	
	\bibitem{Vella2005}
	\bibinfo{author}{Vella, D.} \& \bibinfo{author}{Mahadevan, L.}
	\newblock \bibinfo{title}{{The “Cheerios effect”}}.
	\newblock \emph{\bibinfo{journal}{Am. J. Phys.}}
	\textbf{\bibinfo{volume}{73}}, \bibinfo{pages}{817} (\bibinfo{year}{2005}).

	\bibitem{Moy1994}
	\bibinfo{author}{Moy, V.~T.}, \bibinfo{author}{Florin, E.-L.} \&
	\bibinfo{author}{Gaub, H.~E.}
	\newblock \bibinfo{title}{{Intermolecular forces and energies between ligands
			and receptors}}.
	\newblock \emph{\bibinfo{journal}{Science}} \textbf{\bibinfo{volume}{266}},
	\bibinfo{pages}{257--259} (\bibinfo{year}{1994}).
	
	\bibitem{Bahrami2014a}
	\bibinfo{author}{Bahrami, A.~H.} \emph{et~al.}
	\newblock \bibinfo{title}{{Wrapping of nanoparticles by membranes}}.
	\newblock \emph{\bibinfo{journal}{Adv. Coll. Interfac.}}
	\textbf{\bibinfo{volume}{208}}, \bibinfo{pages}{214--224}
	(\bibinfo{year}{2014}).

	\bibitem{Helfrich1973}
	\bibinfo{author}{Helfrich, W.}
	\newblock \bibinfo{title}{{Elastic properties of lipid bilayers: theory and
			possible experiments}}.
	\newblock \emph{\bibinfo{journal}{Z. Naturforsch. C}}
	\textbf{\bibinfo{volume}{28}}, \bibinfo{pages}{693--703}
	(\bibinfo{year}{1973}).
	
	\bibitem{Pecreaux2004}
	\bibinfo{author}{P{\'{e}}cr{\'{e}}aux, J.}, \bibinfo{author}{D{\"{o}}bereiner,
		H.~G.}, \bibinfo{author}{Prost, J.}, \bibinfo{author}{Joanny, J.~F.} \&
	\bibinfo{author}{Bassereau, P.}
	\newblock \bibinfo{title}{{Refined contour analysis of giant unilamellar
			vesicles}}.
	\newblock \emph{\bibinfo{journal}{Eur. Phys. J. E}}
	\textbf{\bibinfo{volume}{13}}, \bibinfo{pages}{277--290}
	(\bibinfo{year}{2004}).
	
	\bibitem{Rawicz2000}
	\bibinfo{author}{Rawicz, W.}, \bibinfo{author}{Olbrich, K.~C.},
	\bibinfo{author}{McIntosh, T.}, \bibinfo{author}{Needham, D.} \&
	\bibinfo{author}{Evans, E.}
	\newblock \bibinfo{title}{{Effect of chain length and unsaturation on
			elasticity of lipid bilayers.}}
	\newblock \emph{\bibinfo{journal}{Biophysical J.}}
	\textbf{\bibinfo{volume}{79}}, \bibinfo{pages}{328--339}
	(\bibinfo{year}{2000}).

	\bibitem{Lau2009}
	\bibinfo{author}{Lau, A. W.~C.}, \bibinfo{author}{Prasad, A.} \&
	\bibinfo{author}{Dogic, Z.}
	\newblock \bibinfo{title}{{Condensation of isolated semi-flexible filaments
			driven by depletion interactions}}.
	\newblock \emph{\bibinfo{journal}{Europhys. Lett.}}
	\textbf{\bibinfo{volume}{87}}, \bibinfo{pages}{48006} (\bibinfo{year}{2009}).
	
	\bibitem{Upadhyayula2012}
	\bibinfo{author}{Upadhyayula, S.} \emph{et~al.}
	\newblock \bibinfo{title}{{Coatings of polyethylene glycol for suppressing
			adhesion between solid microspheres and flat surfaces}}.
	\newblock \emph{\bibinfo{journal}{Langmuir}} \textbf{\bibinfo{volume}{28}},
	\bibinfo{pages}{5059--5069} (\bibinfo{year}{2012}).
	
	\bibitem{Crocker1994}
	\bibinfo{author}{Crocker, J.~C.} \& \bibinfo{author}{Grier, D.~G.}
	\newblock \bibinfo{title}{{Microscopic measurement of the pair interaction
			potential of charge-stabilized colloid}}.
	\newblock \emph{\bibinfo{journal}{Phys. Rev. Lett.}}
	\textbf{\bibinfo{volume}{73}}, \bibinfo{pages}{352--355}
	(\bibinfo{year}{1994}).
	
	\bibitem{Tamba2011}
	\bibinfo{author}{Tamba, Y.}, \bibinfo{author}{Terashima, H.} \&
	\bibinfo{author}{Yamazaki, M.}
	\newblock \bibinfo{title}{{A membrane filtering method for the purification of
			giant unilamellar vesicles}}.
	\newblock \emph{\bibinfo{journal}{Chem. Phys. Lipids}}
	\textbf{\bibinfo{volume}{164}}, \bibinfo{pages}{351--358}
	(\bibinfo{year}{2011}).

	\bibitem{Binks2006a}
	\bibinfo{author}{Binks, B.~P.} \& \bibinfo{author}{Horozov, T.~S.}
	\newblock \emph{\bibinfo{title}{{Colloidal particles at liquid interfaces}}}
	(\bibinfo{publisher}{Cambridge University Press}, \bibinfo{year}{2008}).

	\bibitem{Sharifi-mood2015}
	\bibinfo{author}{Sharifi-mood, N.}, \bibinfo{author}{Liu, I.B.} \&
	\bibinfo{author}{Stebe, K.J.}
	\newblock \bibinfo{title}{{Curvature capillary migration of microspheres}}.
	\newblock \emph{\bibinfo{journal}{Soft Matter}}
	\textbf{\bibinfo{volume}{11}}, \bibinfo{pages}{6768-6779}
	(\bibinfo{year}{2015}).
	
	\bibitem{Sarfati2016a}
	\bibinfo{author}{Sarfati, R.} \& \bibinfo{author}{Dufresne, E.R.}
	\newblock \bibinfo{title}{{Long-range attraction of particles adhered to lipid vesicles}}.
	\newblock \emph{\bibinfo{journal}{Phys. Rev. E}}
	\textbf{\bibinfo{volume}{94}}, \bibinfo{pages}{012604}
	(\bibinfo{year}{2016}).

	\bibitem{Lin2011a}
	\bibinfo{author}{Lin, H.~K.}, \bibinfo{author}{Zandi, R.},
	\bibinfo{author}{Mohideen, U.} \& \bibinfo{author}{Pryadko, L.~P.}
	\newblock \bibinfo{title}{{Fluctuation-induced forces between inclusions in a
			fluid membrane under tension}}.
	\newblock \emph{\bibinfo{journal}{Phys. Rev. Lett.}}
	\textbf{\bibinfo{volume}{107}}, \bibinfo{pages}{2--6} (\bibinfo{year}{2011}).
	
	\bibitem{Reynwar2011}
	\bibinfo{author}{Reynwar, B.~J.} \& \bibinfo{author}{Deserno, M.}
	\newblock \bibinfo{title}{{Membrane-mediated interactions between circular
			particles in the strongly curved regime}}.
	\newblock \emph{\bibinfo{journal}{Soft Matter}} \textbf{\bibinfo{volume}{7}},
	\bibinfo{pages}{8567} (\bibinfo{year}{2011}).
	
	\bibitem{Saric2012}
	\bibinfo{author}{{\v{S}}ari{\'{c}}, A.} \& \bibinfo{author}{Cacciuto, A.}
	\newblock \bibinfo{title}{{Mechanism of membrane tube formation induced by
			adhesive nanocomponents}}.
	\newblock \emph{\bibinfo{journal}{Phys. Rev. Lett.}}
	\textbf{\bibinfo{volume}{109}}, \bibinfo{pages}{188101}
	(\bibinfo{year}{2012}).
	
	\bibitem{dan1993}
	\bibinfo{author}{Dan, N.}, \bibinfo{author}{Pincus, P.} \&
	\bibinfo{author}{Safran, S.A.}
	\newblock \bibinfo{title}{{Membrane-induced interactions between inclusions}}.
	\newblock \emph{\bibinfo{journal}{Langmuir}}
	\textbf{\bibinfo{volume}{9}}, \bibinfo{pages}{2768--2771}
	(\bibinfo{year}{1993}).
	
	\bibitem{Callan-Jones2011}
	\bibinfo{author}{Callan-Jones, A.}, \bibinfo{author}{Sorre, B.} \& \bibinfo{author}{Bassereau, P.}
	\newblock \bibinfo{title}{{Curvature-driven lipid sorting in biomembranes}}.
	\newblock \emph{\bibinfo{journal}{Cold Spring Harb. Perspect. Biol.}}
	\textbf{\bibinfo{volume}{3}}, \bibinfo{pages}{1--14}
	(\bibinfo{year}{2011}).

	\bibitem{Katira2015}
	\bibinfo{author}{Katira, S.}, \bibinfo{author}{Mandadapu, K.~K.}
	\bibinfo{author}{Vaikuntanathan, S.}, \bibinfo{author}{Smit, B.} \&
	\bibinfo{author}{Chandler, D.}
	\newblock \bibinfo{title}{{Pre-transition effects mediate forces of assembly between transmembrane proteins}}.
	\newblock \emph{\bibinfo{journal}{arXiv}} \bibinfo{pages}{1506.04298}
	(\bibinfo{year}{2015}).

	\bibitem{Appel}
	\bibinfo{author}{Appel, J.}, \bibinfo{author}{Akerboom, S.},
	\bibinfo{author}{Fokkink, R.~G.} \& \bibinfo{author}{Sprakel, J.}
	\newblock \bibinfo{title}{{Facile one-step synthesis of monodisperse
			micron-sized latex particles with highly carboxylated surfaces}}.
	\newblock \emph{\bibinfo{journal}{Macromol. Rapid Comm.}}
	\textbf{\bibinfo{volume}{34}}, \bibinfo{pages}{1284--1288}
	(\bibinfo{year}{2013}).
	
	\bibitem{Meng2004}
	\bibinfo{author}{Meng, F.}, \bibinfo{author}{Engbers, G. H.~M.} \&
	\bibinfo{author}{Feijen, J.}
	\newblock \bibinfo{title}{{Polyethylene glycol-grafted polystyrene particles}}.
	\newblock \emph{\bibinfo{journal}{J. Biomed. Mater. Res. A}} \textbf{\bibinfo{volume}{70}}, \bibinfo{pages}{49--58}
	(\bibinfo{year}{2004}).
	
	\bibitem{Angelova1986a}
	\bibinfo{author}{Angelova, M.~I.} \& \bibinfo{author}{Dimitrov, D.~S.}
	\newblock \bibinfo{title}{{Liposome electroformation}}.
	\newblock \emph{\bibinfo{journal}{Faraday Discuss.}}
	\textbf{\bibinfo{volume}{81}}, \bibinfo{pages}{303--311}
	(\bibinfo{year}{1986}).
	
	\bibitem{Crocker1996}
	\bibinfo{author}{Crocker, J.~C.} \& \bibinfo{author}{Grier, D.~G.}
	\newblock \bibinfo{title}{{Methods of digital video microscopy for colloidal
			studies}}.
	\newblock \emph{\bibinfo{journal}{J. Colloid Interf. Sci.}}
	\textbf{\bibinfo{volume}{179}}, \bibinfo{pages}{298--310}
	(\bibinfo{year}{1996}).
	
	\bibitem{Allan2015}
	\bibinfo{author}{Allan, D.}, \bibinfo{author}{Caswell, T.},
	\bibinfo{author}{Keim, N.} \& \bibinfo{author}{van~der Wel, C.}
	\newblock \bibinfo{title}{{Trackpy v0.3.0}}.
	\newblock \emph{\bibinfo{journal}{Zenodo}}
	\bibinfo{pages}{34028} (\bibinfo{year}{2015}).
	\newblock \urlprefix\url{http://zenodo.org/record/34028}.
	
	\bibitem{VanderWel2016}
	\bibinfo{author}{van~der Wel, C.}
	\newblock \bibinfo{title}{{Circletracking v1.0}}.
	\newblock \emph{\bibinfo{journal}{Zenodo}}
	\bibinfo{pages}{47216} (\bibinfo{year}{2016}).
	\newblock \urlprefix\url{https://zenodo.org/record/47216}.
	
\end{thebibliography}
\end{document}